\documentclass{IEEEcsmag}

\usepackage[dvipsnames]{xcolor}
\usepackage[colorlinks,urlcolor=MidnightBlue,linkcolor=MidnightBlue,citecolor=MidnightBlue]{hyperref}

\jvol{}
\jnum{}
\paper{}
\jmonth{}
\jname{To Appear in IEEE Security \& Privacy: Special Issue on Usable Security for Security Workers}
\pubyear{2023}

\usepackage{subcaption}

\begin{document}

\sptitle{}
\editor{}

\title{Embedding Privacy Into Design Through Software Developers: Challenges \& Solutions}

\author{Mohammad Tahaei}
\affil{University of Bristol}

\author{Kami Vaniea}
\affil{University of Edinburgh}

\author{Awais Rashid}
\affil{University of Bristol}

\markboth{}{}

\begin{abstract}
To make privacy a first-class citizen in software, we argue for equipping developers with usable tools, as well as providing support from organizations, educators, and regulators. We discuss the challenges with the successful integration of privacy features and propose solutions for stakeholders to help developers perform privacy-related tasks.
\end{abstract}

\maketitle

\chapterinitial{Software developers} have become an indispensable part of the digital data-driven economy, with their products contributing to all aspects of society. However, privacy issues arising from using this software are not fully considered when developing software systems. Perhaps thirty years ago, it was enough to have a password to limit access to a database, but in today's software development, developers\footnote{Within the context of this article, we broadly use the term \textit{developer} to refer to anyone who writes computer programs. A developer in the modern software development ecosystem can range from a hobbyist or someone who had only one course in programming to someone with many years of experience in a specific programming language~\cite{tahaei2022recruiting}. Interested readers can find the details of the population's demographics and experience in the cited papers.} need to consider various other aspects to protect users' privacy and comply with privacy laws---example scenarios are: how to store users' data safely and privately, put in place procedures for future deletion of data on user request, ask for users' permission when requesting to read a sensor, or think twice before collecting data to ensure that whatever is collected is needed to avoid the burden of tracking and deleting that specific data in the future.

The privacy infringements on users' data that now and then come out in the news, such as using Facebook data for political campaigns~\cite{lepochat2022audit}, collecting sensitive data from child-directed apps~\cite{reyes2018wont}, and leaking of personal data because of using mobile analytics services, show that privacy is not yet addressed effectively in software companies, regardless of their size. Developers in a large company may have access to dedicated privacy staff to address privacy-related issues, which may cause clashes between teams with diverging priorities (e.g., functionality vs. privacy). On the other hand, those who work in small and medium-sized companies may not have access to experts' advice about privacy or costly tools for testing privacy features. Therefore, it is essential to acknowledge that developers have various backgrounds, contexts, expertise, and individual concerns and characteristics that can impact their programming and priorities~\cite{tahaei2021champions,tahaei2020so,tahaei2019stast}.

Modern software development entails writing components and building complex programs that both use and are used by others. In other words, modern developers do not entirely write their software from scratch; instead, they often use libraries and application programming interfaces written by other developers. Consequently, a modern developer is unlikely to fully understand everything the resulting software does. This situation has already been extensively discussed in terms of security, where a vulnerability in a commonly used library means that all software using it is also vulnerable~\cite{tahaei2019survey}. 

In terms of privacy, it means that the privacy implementation decisions of one developer can directly impact the privacy behavior of all software that uses that component, often in hard-to-detect ways. A typical example is advertising libraries that may collect private data, such as location, without the developer's knowledge~\cite{tahaei2022charting}. A less obvious example is error logs where a third-party library may collect crash reports containing personal information~\cite{zhang2020how}. Such data collection can even impact children; 19\% of the most popular child-directed Android apps collected sensitive data such as identifiers, location data, and email from users, which is against the law (data was collected from 2016 to 2018)~\cite{reyes2018wont}. Often, data collection occurs without the developer's knowledge through included libraries due to modern complex software development~\cite{reardon201950ways}.

Even when a developer wants to preserve their users' privacy, the software ecosystem's complexity and the stakeholders' differing goals can make it challenging~\cite{tahaei2022charting, tahaei2021deciding}. Returning to the example of an advertising library, many such libraries offer developers the ability to limit data collection, but finding the settings may involve going through multiple layers of settings or looking up what sample code function parameters actually do~\cite{tahaei2021what,tahaei2022charting}. Given the challenges, it is unsurprising that some developers opt to use more straightforward approaches such as feeding libraries random numbers instead of potentially sensitive data or finding ways to avoid using software from larger companies perceived as privacy-unfriendly~\cite{tahaei2020so,tahaei2022advice}. 

Privacy and the General Data Protection Regulation (GDPR) issues may naturally fall in the domains of law and public policy, so one might argue that developers should not need to get involved with these topics. We argue that currently, developers have no choice because modern software development is intertwined with handling people's (sensitive) data, requiring developers to make many small code-level decisions that can significantly impact users' privacy. As a result, developers need an understanding of how their code decisions impact users' privacy. They also need access to usable tools that help them understand the issues and make informed decisions. Such approaches could provide developers with direct support in addition to the support that may be provided at higher levels, such as having a data protection officer whom they can ask for advice.

This article argues that one way to improve privacy in the digital economy is through its developers. We reflect on four years of our own research (2019--2022) related to privacy with a specific focus on developers to provide a \textit{synthesis} and a \textit{secondary analysis} of our findings.\footnote{Throughout when a citation is included in a statement or paragraph; it is based on prior published work. Otherwise, it is an interpretation, an assumption, or a post-publication thought.}

Our research suggests that implementing privacy as a transdisciplinary topic is hard for developers because software development platforms do not offer usable support to developers. Privacy is often treated as a marginalized secondary feature in developers' tools that influence their priorities and cause privacy to become an afterthought. Our findings show that organizational factors such as dedicating time and staff can improve the privacy posture of companies. We further discuss educators' and regulators' influence on the privacy understanding of developers and suggest solutions to support developers in performing privacy tasks.

\section{How Did We Come Up With the Insights?}
We employed several qualitative and quantitative methods to understand what makes privacy difficult for a developer and how we can make it easier. Here, we briefly summarize our research methodology over the past four years:

We \textit{interviewed} 12 privacy champions (i.e., people who strongly care about advocating privacy) to understand their motivations, strategies, and challenges when promoting privacy in software teams and organizations~\cite{tahaei2021champions}. We also interviewed 20 computer science students to determine whether they would consider privacy features when designing an app~\cite{tahaei2019stast}. While small in the number of interviewees, these studies gave us a thorough understanding of our participants' thought processes, decision-making, and practices.

In a \textit{survey} with 400 participants with mobile app development experience, we evaluated the impact of design in privacy-related interfaces on developers' choices and the possibility of nudging them to make a privacy-friendly decision~\cite{tahaei2021deciding}. We further assessed developers' understanding and ability to find privacy-related interfaces on mobile advertising networks by running 11 \textit{think-aloud} sessions with developers, two usability experts, and several in-depth sessions with a senior developer, which gave us a chance to extensively analyze the usability of privacy interfaces directed at developers~\cite{tahaei2022charting,tahaei2021what,tahaei2021codeLevel}.

To further triangulate our data points and dig deeper into developers' privacy practices, in three studies, we analyzed a large set of privacy-related posts from \textit{Stack~Overflow} to gather viewpoints of a large pool of developers around privacy-related topics~\cite{tahaei2020so, tahaei2022advice, tahaei2022exploration}. The results gave us an understanding of developers' privacy challenges and solutions in developers' words by analyzing their posts about privacy.

\section{Challenges and Solutions}
\begin{figure*}
    \centering
    \includegraphics[width=\textwidth]{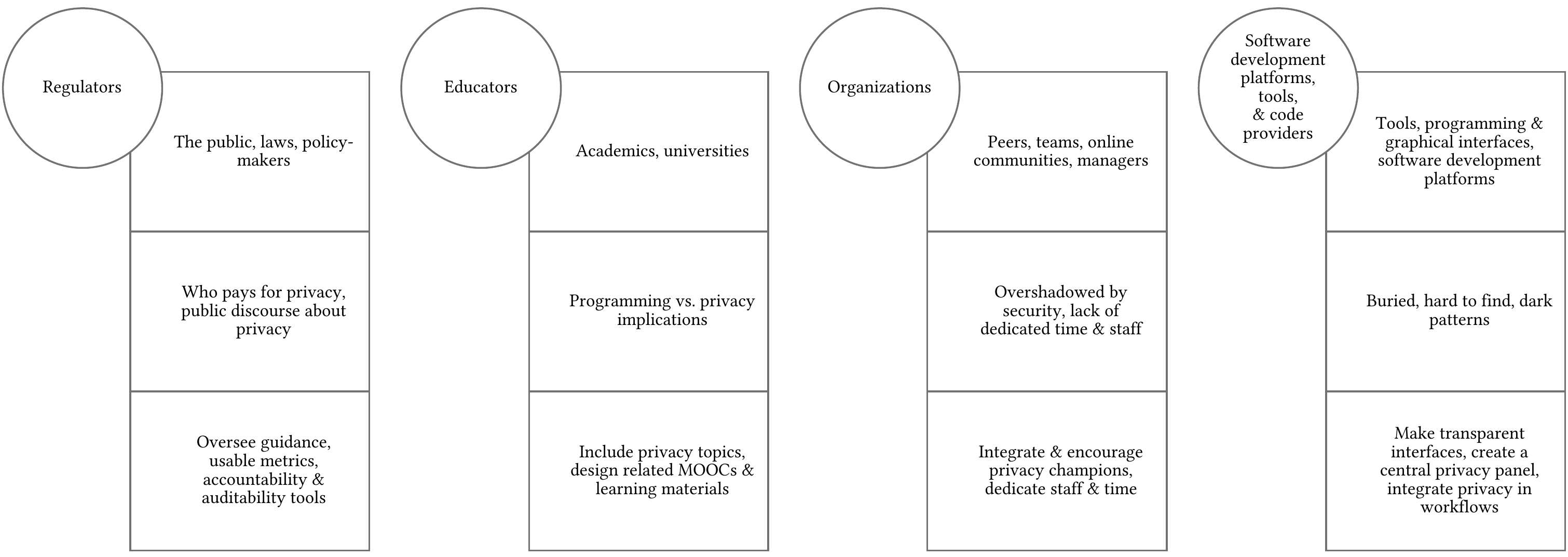}
    \caption{An overview of our findings. The circles show the four identified stakeholders. The three boxes represent involved entities, challenges, and solutions.}
    \label{fig:overview}
\end{figure*}

Our research suggests that privacy features broadly create \textit{hurdles} for developers because they require the engagement of multiple parties, are closely associated with laws, and entail engineering \textit{challenges}. Figure~\ref{fig:overview} gives an overview of the involved stakeholders in the software development ecosystem, the associated challenges, and the potential \textit{solutions} to smooth the hurdles. In the following sections, we elaborate on this figure's entities.

\subsection{Software Development Platforms, Tools, and Code Providers}
\label{sec:tools}
Software development platforms, tools, and code providers directly interact with developers. Examples include a third-party library to log users' interactions and crashes, as well as ad libraries that help developers monetize their mobile apps with graphical interfaces to choose between personalized and non-personalized ads.

\subsubsection{Challenges.}
Software development tools may not put privacy first and focus on revenue and functionality, which causes privacy to become a \textit{marginalized feature} in such tools. The marginalization creates a hurdle developers must overcome to integrate privacy features. For those not currently thinking about privacy, it also contributes to the invisibility of privacy since nothing in their development toolbox will actively prompt them to think about the issue~\cite{tahaei2022charting}.

From a user interface perspective, \textit{navigating through privacy interfaces} in software development platforms directed at developers (e.g., developer panels in advertising networks and programming libraries) can be time-consuming, frustrating, and confusing. Some interfaces bury privacy settings under several layers. Such treatment and where privacy is located on the interfaces hints at the level of attention privacy will likely get from developers. For example, forcing a developer to go down five levels into an interface means that only a developer actively looking for a privacy control to disable personalized ads will find the checkbox~\cite{tahaei2022charting, tahaei2021what, tahaei2021codeLevel}.

These interfaces may even contain \textit{dark patterns} (i.e., design patterns meant to nudge users into making a decision in the platform owner's best interest instead of the user's) to manipulate developers' choices. While the impact of a dark pattern on an end-user may only influence that individual, in the case of developers, their choices may impact all their users. These decisions can have an enormous impact on users of apps, and it puts the burden of privacy on the developers' shoulders instead of the platforms'. Developers may not read all the policies or terms of service and consequently not be aware that they bear the legal consequences associated with using third-party platforms rather than the platforms themselves~\cite{tahaei2022charting,tahaei2021what}. 

\begin{figure*}
     \centering
     \begin{subfigure}{\textwidth}
         \centering
         \includegraphics[width=\textwidth]{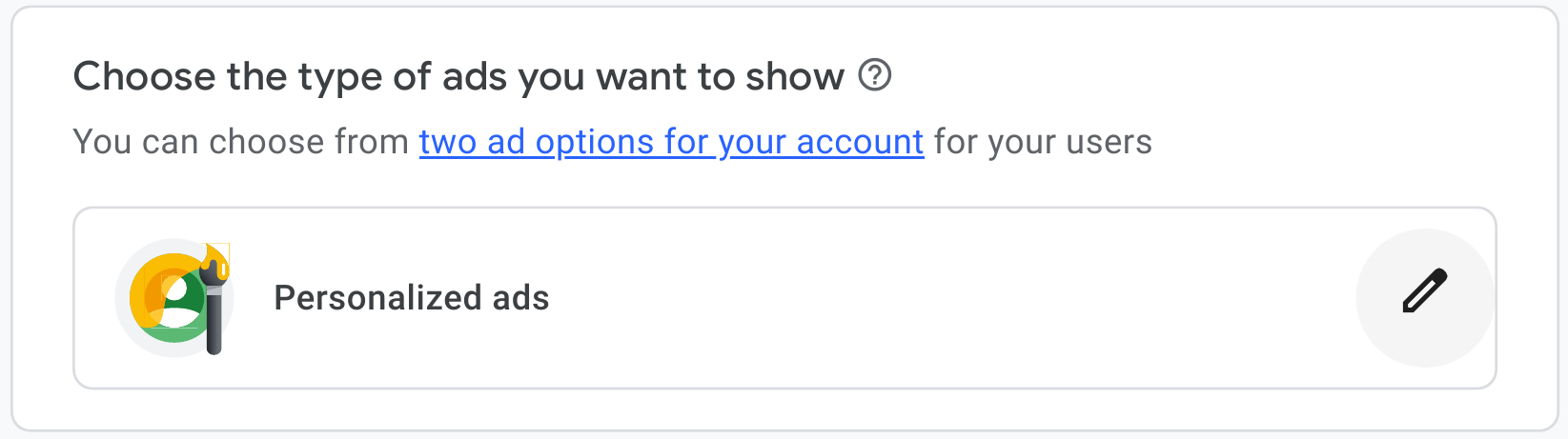}
         \caption{The initial view.}
     \end{subfigure}
     \hfill
     \begin{subfigure}{\textwidth}
         \centering
         \includegraphics[width=\textwidth]{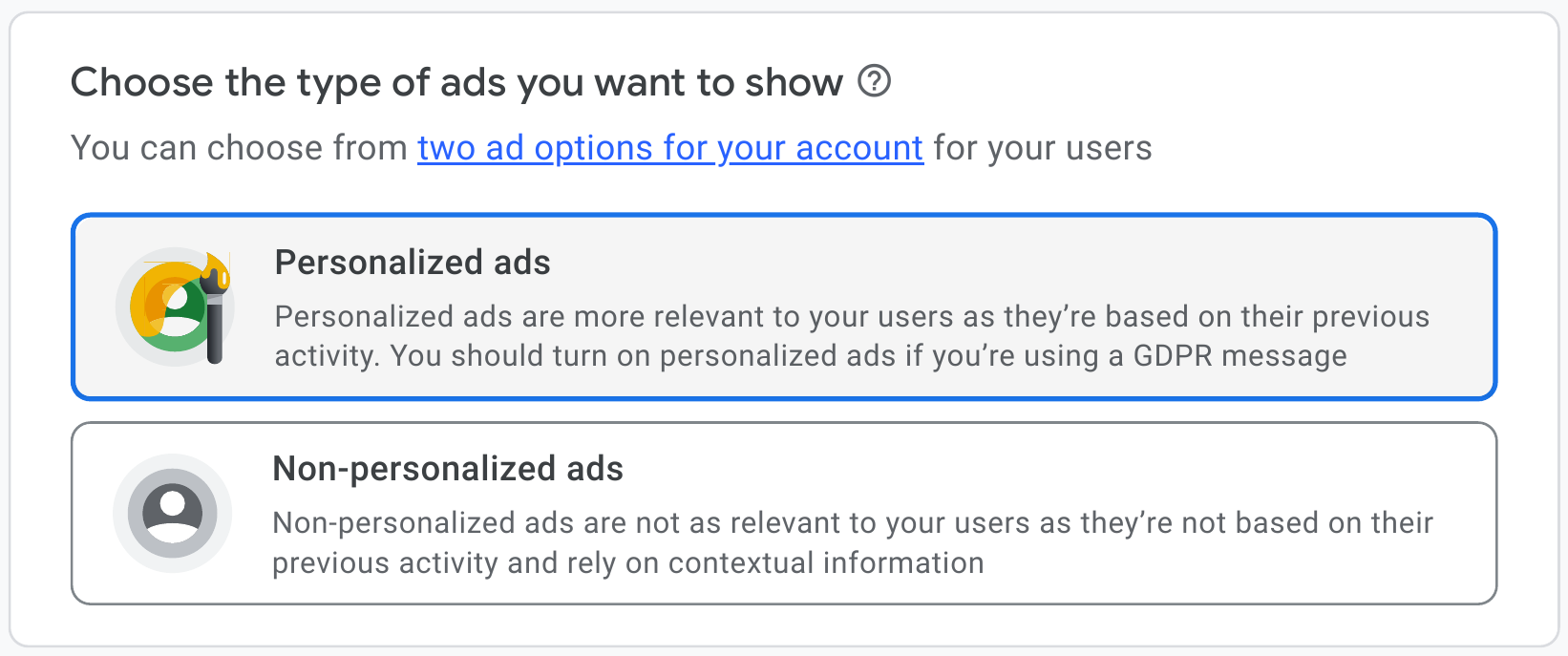}
         \caption{After clicking on the personalized ads on the first row.}
     \end{subfigure}
     \caption{The GDPR section of Google AdMob's developer panel contains several dark patterns: (1) the first option is pre-selected by default to show personalized ads, which means it requires extra effort and time to find these options to show non-personalized ads, (2) the second option in is not visible in the initial view, meaning that the user may not know that there is a second option unless they click on the first option, and (3) the color of the second option stays gray even after selecting and saving the non-personalized option which may hint a lower priority compared to the personalized option and come off less engaging and appealing. Clicking on the ``two options for your account'' opens a new page to explain the two options instead of showing them. Non-personalized ads are generally associated with less data collection. Screenshots were taken in March 2022 with a U.K. IP address.}
    \label{fig:google-gdpr}
\end{figure*}

Figure~\ref{fig:google-gdpr} shows an example from Google AdMob's developer panel for GDPR compliance. It shows that privacy options are hidden under several layers and contain dark patterns to \textit{nudge} developers into choosing a privacy-unfriendly option. Overall, most of the design choices made by the studied platforms show a trade-off between their primary focus (i.e., integrating an ad) and what is an optional choice or direction for developers (e.g., privacy features)~\cite{tahaei2022charting,tahaei2021what,tahaei2021codeLevel}.

The \textit{documentation} of software development platforms (e.g., advertising libraries) is also filled with links and references to policy documents that developers are less likely to read or follow. Those documents explain that developers are the ones who must protect users' privacy and ensure compliance with laws. However, they do not explain how to do so~\cite{tahaei2022charting,tahaei2021what,tahaei2021codeLevel}. Software development platforms might not have had enough time to keep up with the laws and need more time to build usable materials for those new legal additions, which again goes back to the trade-off between the most crucial aspect from their viewpoint compared to what is an optional feature to develop (e.g., functional vs. privacy-preserving features).

Privacy laws commonly manifest in software systems as terms and conditions, and \textit{privacy policies}---notoriously hard-to-read documents. These documents are hard to deal with for users and developers alike. Mobile app stores require app developers to include a privacy policy. Yet, they do not provide information about how to write a privacy policy or what needs to be included. Developers are, therefore, often left with the need to comply with regulations and requirements of the app stores but without the knowledge of what needs to go into the privacy policies~\cite{tahaei2020so,tahaei2022advice}. 

\begin{figure*}
\centering
  \includegraphics[width=\textwidth]{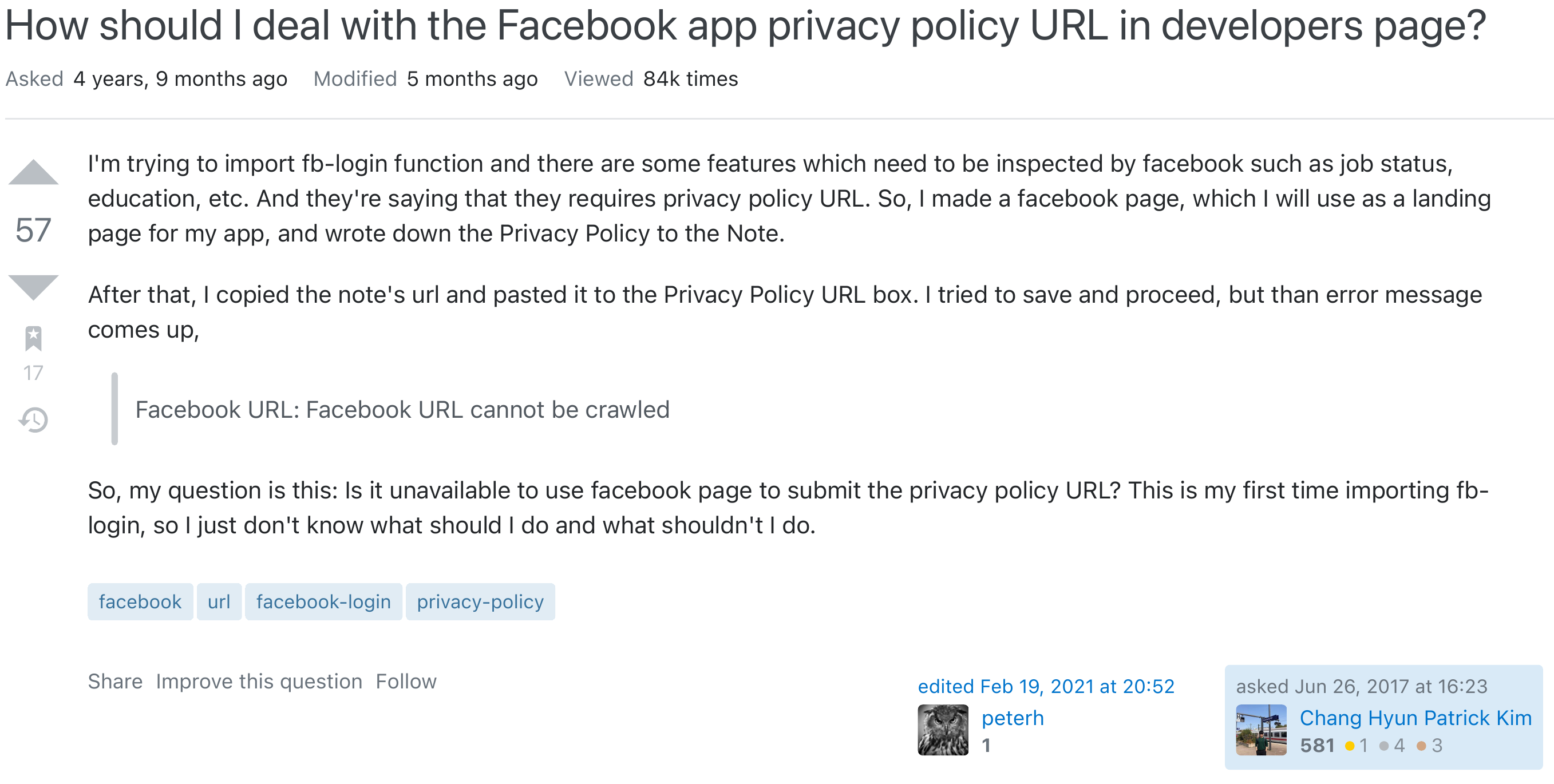}
  \caption{A sample privacy-related question about privacy policies on Stack~Overflow [\href{https://stackoverflow.com/questions/44764212}{stackoverflow.com/44764212}]. The screenshot was taken in March 2022 with a U.K. IP address.}
    \label{fig:sample-question}
\end{figure*}

Figure~\ref{fig:sample-question} shows a sample question on Stack~Overflow about how to include a privacy policy for a Facebook app. Writing these documents may become even more challenging in cases where developers working in small teams might not have access to dedicated privacy experts or lawyers because of cost issues. One consequence for users is privacy policies that are hard to read, may not even talk about the app's privacy policies, and may lack several essential pieces of information~\cite{tahaei2020so, tahaei2022advice}.

\subsubsection{Solutions.}
Our findings suggest that developers with privacy concerns for their users still struggle because they do not have the right tools and skills to address those concerns appropriately. Some developers may also not consider privacy when programming. Either way, we see \textit{software development platforms} as critical players for creating usable and responsibly-designed privacy tools for developers and bringing privacy to developers' attention. However, the current design of platforms marginalizes privacy. Platforms also drive what sensitive data means; if they include location data as sensitive data, developers have to take additional steps to access that resource and consequently learn to take extra care. Such features and requests by platforms resulted in many Stack~Overflow posts about privacy~\cite{tahaei2020so,tahaei2022advice}.

\begin{figure*}
\centering
  \includegraphics[width=\textwidth]{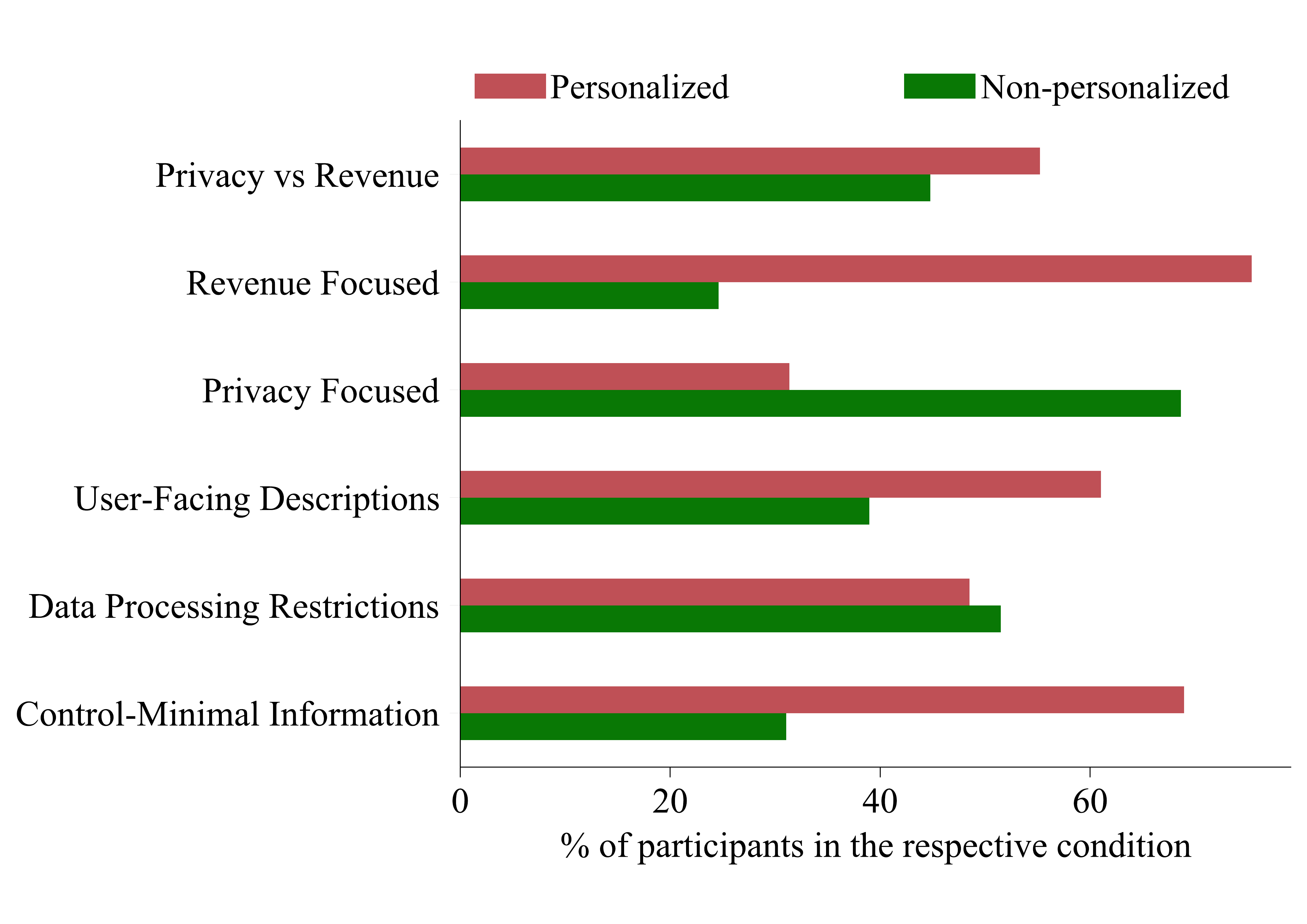}
  \caption{In an experiment with 400 participants with mobile app development experience, we varied the wording for two choices between personalized and non-personalized ads. This plot shows their choices across the six conditions. When participants had minimal information about the privacy consequences (control condition), which is the typical scenario in ad networks, they often chose to show personalized ads. However, when privacy is highlighted (e.g., by saying that the ad company can show users ads based on their past behavior, such as previous visits to sites or apps or where the user has been), participants were significantly more likely to pick the non-personalized option which is a more privacy-preserving choice compared to participants in the control condition (control condition vs. privacy focused condition). For details of the wordings and the conditions, see~\cite{tahaei2021deciding}.}
    \label{fig:conditions}
\end{figure*}

To study the impact of interfaces provided by platforms, we conducted a study where we made \textit{privacy salient} in the choice between personalized and non-personalized ads (Figure~\ref{fig:conditions}). Participants who had to choose between those two options and were told about the privacy consequences of their choices on the users were less likely to pick the personalized option (i.e., collects more user data). Undoubtedly, developers need support from tools and platforms to overcome challenges of privacy~\cite{tahaei2022charting,tahaei2021deciding}.

On the other hand, we also observed a group of \textit{privacy-conscious developers} in Stack~Overflow worried about using libraries or tools built or supported by large technology companies. They thought those companies might silently collect users' data, and thus, they were looking for alternatives or clear explanations before they started using those tools and libraries~\cite{tahaei2020so,tahaei2022advice}. Therefore, we believe there is an opportunity to better support this group of developers who are actively looking for privacy-friendly options by creating alternative privacy-friendly technologies (e.g., for monetization and analytics).

A recurring pain point for mobile app developers, mentioned on Stack~Overflow, was the challenging task of including and \textit{writing a privacy policy}~\cite{tahaei2020so,tahaei2022advice}. One potential avenue to explore is to provide automated tools for small and medium-sized development teams to facilitate creating privacy policies. As part of a long-term goal, creating privacy policies could be supported by a graphical interface in the app stores instead of a separate activity for developers to start from scratch. In an alternative workflow, developers can define permissions in a file or a graphical interface with their intended use and purpose, and the privacy policy will be created for them.

Privacy laws apply to all products, such as voice assistants, online shopping websites, and mobile apps. Much research has been done to make consent forms and privacy settings usable for the public and introduce laws to ensure those interfaces follow what is required (or show that they do not). However, there has not been much research in the developers' space. We suggest expanding the scope of user-focused research to all types of users. Companies that build these tools and libraries should put privacy first and offer transparent and honest options through mindful design. Privacy scandals have damaged the reputation of several companies, which can be avoided by making privacy a central element and making minor changes such as removing murky and unethical design patterns in privacy-related interfaces, for example, by reducing the prominence of a privacy-unfriendly option using colors and ordering or removing the stress on revenue over other considerations in wordings.

Last but not least, appification has driven a democratization effect that has enabled people from all walks of life to
write and deploy software to others. Consequently, programming should be \textit{accessible} to everyone. Therefore, privacy needs to be \textit{inclusive} for all developers (e.g., based on their experience level). Implementations of privacy features should have this consideration in mind during their design and development process, which as a consequence, can result in ease of privacy integration for developers who use them. For example, creating easy-to-use testing mechanisms and guides to help developers test against and follow regulations without additional costs would be a future work avenue to explore.

\subsection{Organizations}
Organizations dictate policies, tools, and priorities---which impacts how developers think about and integrate privacy features~\cite{tahaei2021champions}. 

Not all organizations have access to privacy experts or lawyers with privacy expertise. Even when a team has such access, the challenge is in the interaction between the legal and technical teams as they talk about privacy using different languages. \textit{Translating} legal requirements into tangible technical requirements has been a long-standing challenge in bringing privacy into software design. These clashes between teams can cause delays in implementing privacy features and create hurdles for developers~\cite{tahaei2021champions}.

\subsubsection{Challenges.}
When there is no \textit{dedicated time and staff} for privacy, it naturally becomes a low, a post-design, or an implementation problem. Examples of features that privacy contends with for attention are security and functionality~\cite{tahaei2021champions}. 

While \textit{security} might be viewed as a non-functional requirement and left aside, privacy might be treated even worse~\cite{tahaei2019stast, tahaei2019survey}. Security can work as an enabler for privacy; conversely, it may also work as a deterrent. When privacy is overshadowed by security, and the language of communication is dominated by security, it is challenging to make privacy a first-class citizen in the design process. Overall, the battle and trade-off between various features and teams in product design can cause privacy to be left behind unless a champion is willing to fight for it~\cite{tahaei2021champions}.

\subsubsection{Solutions.}
Instead of having security teams manage privacy features or combining privacy as part of security features, we suggest dedicating time and staff to privacy. One approach is integrating \textit{privacy champions} into teams to promote privacy. We have found that privacy champions can be the voice of privacy and use their soft and technical skills to help others work with privacy features and consider privacy during design and implementation~\cite{tahaei2021champions}. 

Privacy champions are highly passionate about privacy, show empathy toward users, and may have solid personal attitudes toward privacy in many cases. They also believe that privacy could be a competitive advantage for the company, and improving privacy could result in stronger branding and sales~\cite{tahaei2021champions}.

Privacy champions improve a company's privacy culture by discussing privacy topics with other employers, which may not be part of a formal workshop but rather during informal water-cooler chats or contributing to company-wide Slack channels when a privacy-related topic is discussed. Another significant contribution of these individuals is facilitating the privacy conversation between different teams. As stated above, different teams may have different priorities resulting in privacy being marginalized. However, privacy champions can bring teams together to discuss privacy topics~\cite{tahaei2021champions}. 

To support privacy champions, managers can motivate them by \textit{acknowledging their efforts} through financial incentives, verbally giving credit to their work, or helping them stand out in the team with a t-shirt. Companies can make privacy part of their branding and motto to attract and retain privacy champions. Privacy champions are attracted to (and stay in) companies that care about privacy and believe in privacy and may leave companies when privacy is treated poorly~\cite{tahaei2021champions}.

Another space for improvement is building standardized privacy protocols that can help champions argue for privacy with formal policies. This could come from academic work as some of the champions are closely connected with the latest academic research or from regulatory support~\cite{tahaei2021champions}.

We suggest improving organizational privacy through publishing \textit{case studies}. While news about privacy breaches comes out now and then, detailed case studies about companies' privacy best practices and pitfalls are not published as much. Companies may not want to discuss their privacy breaches and practices publicly; however, knowing the problems, best practices, and what has been tried is hard without reading others' lessons learned and case studies. Therefore, we make a call to organizations to publish case studies about their privacy practices.

\subsection{Educators}
Another hurdle that makes it difficult for developers to consider privacy features is their educational background. While some developers have a \textit{computer science-related degree}, their education may not emphasize privacy features or ethical and responsible design~\cite{tahaei2019stast}.

In a study with twenty computer science students, we asked participants to tell us how they would go about designing a simple mobile app. While many participants previously took a security course, security features still did not appear in their design. Such an emphasis on functionality may also lead to developers who emphasize \textit{efficiency} and overlook \textit{responsible design}~\cite{tahaei2019stast}.\footnote{The study covered students in one university in the U.K. with limited background in security and programming. A majority did not have professional development experience. Therefore, the results should be interpreted with these limitations but show preliminary findings. More data points are needed to understand mentioned factors.}

\subsubsection{Challenges.}
A computer science curriculum focuses on software development, mathematics, and algorithms, and less attention is typically given to ethics and privacy. Such a trade-off trains software developers who are competent in developing functional software but may pay less attention to building privacy-preserving systems.

\subsubsection{Solutions.}
On the upside, some software developers have a computer science background, so there is an opportunity to make such developers aware of the privacy consequences of their choices on their users. 

In addition to solutions such as including courses that cover privacy, ethics, and responsible design, we suggest creating and sharing \textit{free lesson plans}, activities, and materials to support academics to include privacy topics in their courses with minimal effort.

The coverage of privacy topics in non-traditional software development learning routes (e.g., coding bootcamps and online courses) and how to best support developers who learn from these resources remain an open area for future research. We suggest academics create materials to reach out to the \textit{developer communities}. While there are thousands of resources, such as blogs, videos, and sample codes, to help developers build an app, there are few resources to help them integrate privacy features. In particular, videos and materials to teach and educate them about privacy and how to do it are currently missed in the documentation of tools and libraries~\cite{tahaei2022charting}. Academics can fill these gaps by offering Massive Open Online Courses (MOOCs) or YouTube videos that specifically cover regulatory and privacy topics of software development.

\subsection{Regulators}
Regulators and policymakers sit with the power to make high-level shifts in how privacy is treated by technologies, companies, educators, and developers.

Privacy as a human right is connected to laws and regulations, and it is a vague multi-faceted term whose interpretation depends on the culture and the individual. When privacy enters the digital realm, it is even more ambiguous because the technologies we use in our everyday lives are not well-established, and we do not yet know their capabilities and \textit{social impacts}.

Privacy in the digital world is a nascent and ever-changing area, meaning related laws, such as the California Consumer Privacy Act (CCPA) and the GDPR, also evolve with the changes in these technologies. Innovative tools nowadays collect data from all locations a person has been to, track their habits, and predict what they want to buy next. For example, changes in traditional tools (e.g., smart home appliances, virtual reality, and wearables) can have privacy consequences, and changes in users' demands and needs can introduce new laws. Similar to the rooted identity of privacy in cultures and individual values, privacy laws reflect their originated countries and are \textit{contextual} and applicable to specific regions.

All these changes and new aspects impose a \textit{burden} on developers. They must adhere to laws from various countries, even if they are not located there, and have users across multiple countries, which means that they need to follow up with several different regulations and laws to make their apps compliant with them~\cite{tahaei2022charting,tahaei2022advice}. 

New laws and changes to previous ones would result in platforms \textit{introducing new requirements} to their developers. For example, with the introduction of CCPA and GDPR, iOS and Android started to ask developers to build permission dialogues with clear explanations and the inclusion of privacy policies, which led to many questions and confusion among developer communities, such as those we observed on Stack~Overflow~\cite{tahaei2020so, tahaei2022advice,tahaei2022exploration}.

\subsubsection{Challenges.}
Some of the changes in the laws are based on public expectations and discourse. Examples include the ``I have got nothing to hide'' notion that has been a trend in the daily conversations about privacy which is also visible in software teams~\cite{tahaei2021champions}. Another tangible example is monetizing the software using privacy-invasive methods (e.g., personalized ads), which might be rooted in the unclear value of privacy for users. Using a ``free'' service and paying by data, influenced by the public, may also end up in software design~\cite{tahaei2021deciding}. All of these may result in developers paying less attention to privacy.

\subsubsection{Solutions.}
Regulators at the top of the chain may not intend to engage with the details of tool building and instead provide high-level guides about privacy (e.g., CCPA and GDPR). However, \textit{translation} of these guides to technical requirements is often left to organizations and developers~\cite{tahaei2021champions}. One approach to bridge the gap between these parties is to fund independent academic research groups or not-for-profits to build tangible and understandable guides for the developer community.

We foresee a need for \textit{detailed guidance, usable metrics, accountability, and auditability tools} to help developers understand what their apps do and what they can do to minimize privacy consequences. Not all developers intentionally invade users' privacy, and often, they intend to protect users' privacy. However, third-party libraries may be the culprit~\cite{tahaei2020so,tahaei2022advice,tahaei2022exploration}. For such scenarios, a free and easy-to-use tool or detailed guidance is needed. These tools and guidance should come from an independent entity or a joint industry-driven tool to ensure public benefits are accounted for in the design of those tools and guidance. We also suggest platform owners (e.g., Android and iOS) locate options related to privacy laws in a central place to make it easier for developers to find them~\cite{tahaei2022charting}.

\section{Final Thoughts}
We identified software development platforms, organizations, educators, and regulators as the stakeholders impacting how developers perform privacy-related tasks. The involvement of these stakeholders makes privacy a \textit{transdisciplinary} and a \textit{contextual} feature in software design. Competing features such as functionality, security, and privacy create trade-offs that result in hurdles for developers. These frictions can be resolved by making privacy a first-class feature in software systems. This article suggests solutions to facilitate this shift by supporting developers of these systems.

The \textit{developers' toolbox} is a central point for their work. The current trend that puts much of the burden on developers to comply with regulations and follow privacy best practices should instead be placed on platforms and tool builders who can better handle them. They should provide transparent privacy choices and remove dark patterns that can nudge developers to make a privacy-unfriendly option. These options should be moved to a central privacy panel to help developers find them quickly and allow them to track the changes to these options without extra hassle. Lastly, to prioritize privacy for developers, tools should include privacy features within developers' workflow, next to other functional features.

Organizations can directly impact developers' decisions. They can incorporate a \textit{privacy champion} program to support developers in performing privacy tasks and promote a culture that cares about privacy and recognizes responsible design elements. Educators can build \textit{free educational materials} about technical privacy topics and promote awareness within the developer community. They can also benefit from materials built by fellow academics to integrate privacy lessons in their current course planning easily. Regulators can create \textit{detailed guides and tools} to help developers test against and comply with regulations without extra costs. 

A significant shift in societies, such as caring for and protecting privacy, requires the \textit{engagement} of many stakeholders---developers among them. Our four years of research show that by understanding developers' needs around privacy, responsibly designing and building developer-friendly tools, and supporting developers in privacy-related tasks, we can move privacy, a \textit{human right}, one step forward in our societies.

\section{Acknowledgments}
We thank the anonymous reviewers and the editors for constructive feedback that helped improve the paper. We also express our gratitude to our collaborators, Julia Bernd, Konstantin Beznosov, Alisa Frik, Jason I. Hong, Adam Jenkins, Tianshi Li, Kopo M. Ramokapane, Naomi Saphra, and Maria K. Wolters, who helped us gain these findings and insights.

This article was sponsored in part by Microsoft Research through its Ph.D. Scholarship Program, Why Johnny doesn't write secure software? Secure software development by the masses (EPSRC: EP/P011799/2), and REPHRAIN: U.K.'s National Research Centre on Privacy, Harm Reduction and Adversarial Influence Online (EPSRC: EP/V011189/1).

\bibliographystyle{IEEEtran}
\bibliography{bibliography}

\begin{IEEEbiography}{Mohammad Tahaei}{\,} is a senior researcher at the University of Bristol with a Ph.D. in Informatics from the University of Edinburgh. His research is on the human factors of computer systems, responsible design, and usable privacy. His current research studies privacy and security technologies and interfaces directed at software developers. He applies qualitative and quantitative methods for evaluating, designing, and building such technologies to assist software developers in building privacy-friendly and secure systems. Find more about his work at \href{https://mohammad.tahaei.com}{mohammad.tahaei.com}.
\end{IEEEbiography}

\begin{IEEEbiography}{Kami Vaniea}{\,}is a Lecturer of cyber security and privacy at the University of Edinburgh, Edinburgh, the United Kingdom, where she heads the Technology Usability Lab in Privacy and Security (TULiPS). Her research focuses on human factors issues around security and privacy technology, including the needs of developers and system administrators as well as end users. Vaniea received a Ph.D. in computer science from Carnegie Mellon University in Pittsburgh, PA, USA. Contact her at \href{mailto:kvaniea@inf.ed.ac.uk}{kvaniea@inf.ed.ac.uk}.
\end{IEEEbiography}

\begin{IEEEbiography}{Awais Rashid}{\,} is a Professor of cybersecurity at the University of Bristol, Bristol, BS8 1QU, the United Kingdom, where he heads the Bristol Cybersecurity Group. His research is focused on security and privacy in large connected infrastructures, software security, and human factors. Rashid received a Ph.D. in computer science from Lancaster University, U.K. He is the director of the National Research Centre on Privacy, Harm Reduction and Adversarial Influence Online (REPHRAIN) and the EPSRC Centre for Doctoral Training in Cybersecurity: Trust, Identity, Privacy and Security in Large-scale Infrastructures (TIPS-at-Scale). He is the principal investigator and editor-in-chief of the Cyber Security Body of Knowledge (CyBOK). Contact him at \href{mailto:awais.rashid@bristol.ac.uk}{awais.rashid@bristol.ac.uk}.
\end{IEEEbiography}

\end{document}